\documentclass[aps,prb,amsmath,amssymb,twocolumn,longbibliography,superscriptaddress]{revtex4-2}
\usepackage{graphicx}
\usepackage{dcolumn}
\usepackage{bm}
\usepackage{xcolor}
\usepackage{mathrsfs}
\usepackage{subfigure}
\usepackage{amssymb}
\usepackage{multirow}
\usepackage{float}
\usepackage{color}
\usepackage{natbib}
\usepackage{soul}

\begin{document}

\title{Tailoring of the interference-induced surface superconductivity\\ by an applied electric field}

\author{Yunfei Bai}
\affiliation{Key Laboratory of Optical Field Manipulation of Zhejiang Province, Department of Physics, Zhejiang Sci-Tech University, 310018 Zhejiang, China}

\author{Libo Zhang}
\affiliation{College of Physics and Optoelectronic Engineering, Hangzhou Institute for Advanced Study, University of Chinese Academy of Sciences, 310024 Zhejiang, China}

\author{Xiaobing Luo}
\affiliation{Key Laboratory of Optical Field Manipulation of Zhejiang Province, Department of Physics, Zhejiang Sci-Tech University, 310018 Zhejiang, China}

\author{A. A. Shanenko}
\affiliation{HSE University, 101000 Moscow, Russia}

\author{Yajiang Chen}
\email{yjchen@zstu.edu.cn}
\affiliation{Key Laboratory of Optical Field Manipulation of Zhejiang Province, Department of Physics, Zhejiang Sci-Tech University, 310018 Zhejiang, China}

\date{\today}
\begin{abstract}
Nucleation of the pair condensate near surfaces above the upper critical magnetic field and the pair-condensate enhancement/suppression induced by changes in the electron-phonon interaction at interfaces are the most known examples of the surface superconductivity. Recently, another example has been reported, when the surface enhancement of the critical superconducting temperature occurs due to quantum interference. In this case the pair states spread over the entire volume of the system while exhibiting the constructive interference near the surface. In the present work we investigate how an applied electric field impacts the interference-induced surface superconductivity. The study is based on a numerical solution of the self-consistent Bogoliubov-de Gennes equations for a one-dimensional attractive Hubbard model. Our results demonstrate that the surface superconducting characteristics, especially the surface critical temperature, are sensitive to the applied electric field and can be tailored by changing its magnitude.
\end{abstract}

\maketitle

\section{Introduction}
\label{introduction}

The surface superconductivity was first predicted by Saint-James and de Gennes in their classical paper~\cite{saint-james1963} with the linearized Ginzburg-Landau (GL) equations under parallel fields $H_{c2}\leq H\leq H_{c3}$, where $H_{c2}$ and $H_{c3}$ are the upper critical and nucleation magnetic fields, respectively. It was confirmed experimentally on various superconducting metallic alloys~\cite{bonmardion1964, hempstead1964, tomasch1964, gygax1964, strongin1964}. Since then, various properties of the ratio $H_{c3}/H_{c2}$ have been reported, including the influence of temperature~\cite{hu1969, hu1969a, ebner1969}, pairing mechanism~\cite{khlyustikov1987, keller1994, samokhin1995, kogan2002, jang2009}, sample geometry~\cite{moshchalkov1995, fomin1998,schweigert1999}, fluctuation corrections and disorder~\cite{agterberg1996,gorokhov2005,aleiner2015,xie2017}, etc. 

It was also revealed that the surface (interface) superconductivity can be significantly different from bulk one when the phonon properties at surfaces/interfaces are altered as compared to the bulk lattice vibrations. The relevant examples range from thin films to small superconductive particles~\cite{strongin1968, dickey1968,naugle1973,chen2012}. 

Many efforts have been made to search for the surface superconducting state in metals without magnetic fields and surface phonon modes. The surface superconducting pair potential (gap function) can indeed be much stronger (up to $\sim20\%$) than its bulk value~\cite{troy1995, shanenko2006,lauke2018,chen2018}. However, the relative difference [$\tau=(T_{cs}-T_{cb})/T_{cb}$] between the surface superconducting transition temperature $T_{cs}$ and the bulk transition temperature $T_{cb}$ was found to be negligible in those cases ($\sim10^{-3}$)~\cite{giamarchi1990}. Finally, it was recently demonstrated by numerically solving the self-consistent Bogoliubov-de Gennes (BdG) equations for an attractive Hubbard model that $\tau$ can increase up to about $25\%$ at the half-filling level~\cite{barkman2019,samoilenka2020,samoilenka2020a}.
In this case the surface superconductivity has a gaped excitation spectrum~\cite{chen2022}, contrary to that in Ref.~\cite{saint-james1963, gennes1964}. The underlying physics is the constructive interference of the pair states near the sample surface~\cite{croitoru2020}. Moreover, this interference-induced surface superconductivity can be further enhanced by tuning the Debye frequency~\cite{bai2023} due to the removal of the contribution of high-energy quasiparticles. As a result, $\tau$ can be enlarged up to $60\%\sim70\%$. However, this value can be smaller for a more sophisticated variant of the confining potential barrier for charge carriers (as compared to the infinite wall)~\cite{croitoru2023}.

Experimentally and theoretically, it is of great importance to investigate the response of the interference-induced surface superconducting state to other controllable parameters, in particular, to an electric field that is one of the most useful tools to modify properties of thin superconductors and surface superconductivity in bulk samples~\cite{glover1960, meissner1967, shapiro1984,amoretti2022}. For example, an electric-field-induced shift of $T_c$ was observed in Sn, In and NbSe$_2$ thin films~\cite{glover1960,bonfiglioli1962,staley2009}. Moreover, the electric field can also give rise to the multigap structure of the surface pair states~\cite{mizohata2013} and result in the superconductor-metal~\cite{golokolenov2021, paolucci2021}, and superconductor-insulator transitions~\cite{parendo2005, ueno2008,yin2020}.

In the present work, we investigate the effect of an external electric field on the interference-induced surface superconductivity within a one-dimensional attractive Hubbard model at the half-filling level. By numerically solving the self-consistent BdG equations, we demonstrate that varying the field strength makes it possible to fine-tune the surface superconducting properties, changing the surface critical temperature in a wide range of its values.

The paper is organized as follows. In Sec.~\ref{theory} we outline the BdG formalism for a one-dimensional attractive Hubbard model in the presence of a screened electric field parallel to the chain of atoms. Our numerical results and related discussions are presented in Sec.~\ref{results}. Concluding remarks are given in Sec.~\ref{conclusions}.

\section{Theoretical formalism}
\label{theory}

Similarly to the previous papers on the interference-induced surface superconductivity~\cite{croitoru2020,bai2023,croitoru2023}, we investigate an attractive ($s$-wave pairing) Hubbard model for a one-dimensional chain of atoms with the grand-canonical Hamiltonian given by~\cite{samoilenka2020a,tanaka2000,takasan2019}:
 \begin{align}\label{hubbard}
H - \mu N_e  =& -\sum_{i\delta\sigma}t_\delta c^{\dagger}_{i+\delta,\sigma}c_{i\sigma} + \sum_{i\sigma}\big[U(i) -\mu\big] n_{i\sigma}  \nonumber \\ 
&  -g \sum_i n_{i\uparrow}n_{i\downarrow},
\end{align}
where $i$ is the site index, $c_i$ and $c^\dagger_i$ are the electron annihilation and creation operators associated with site $i$, $N_e$ and $n_{i\sigma}$ are the total and local electron number operators, $g$ denotes the on-site attractive interaction ($g>0$), $U(i)$ and $\mu$ are the one-electron and chemical potentials, respectively, and $t_\delta$ is the hopping amplitude between sites $i$ and $i+\delta$. We adopt the nearest neighbouring hopping, i.e. $\delta=\pm1$, and $t_\delta=t$. The open boundary conditions are applied in the present study so that the relevant wavefunctions vanish at $i = 0$ and $N + 1$.

The single-electron potential $U(i)$ is the potential energy of an electron in the external electric field ${\bf E}(i)$ that is parallel to the chain and along its positive direction. The field magnitude is given by
\begin{align}\label{Ei}
E(i) &= E_0[e^{-i/\lambda} + e^{-(N+1-i)/\lambda}]  \nonumber \\
& = 2\,E_{0}\,e^{-(N+1)/2\lambda} \cosh\big[\big(2i-N-1\big)/2\lambda\big].
\end{align}
where $E_0$ is the strength of the screened electric field, and $\lambda$ is the screening length in the units of the lattice constant $a$. From Eq.~(\ref{Ei}) one obtains
\begin{equation}
U(i) = -2q\lambda E_0 e^{-(N+1)/2\lambda}\sinh\big[(2i-N-1)/2\lambda\big],
\end{equation}
with $q=-e$ the electron charge. The determination of $\lambda$ near the system surface is rather complex. However, as we are interested in the qualitative picture, we can assume, for simplicity, that $\lambda$ is proportional to the Fermi wavelength $\lambda_F$ of the system in the absence of the electric field and electron attractive interactions $g$, i.e. $\lambda\approx\gamma\lambda_F$, with $\gamma \sim 1$ the parameter of our calculations. Using the dispersion relation~\cite{tanaka2000} $\xi_k=-2t{\rm cos}(ka)$, with $ka = n\pi/(N+1)$, one concludes that the half-filling case corresponds to $\mu=0$. Then, adopting the parabolic band approximation one gets $\lambda_F=\sqrt{2}\pi a$. Below our results are shown for $\gamma=2$. We remark that our qualitative results are not sensitive to this value.

The BdG equations obtained in the mean-field approximation for the Hamiltonian~Eq. (\ref{hubbard}) can be written as~\cite{chen2022,bai2023},
\begin{equation}\label{bdg}
\begin{aligned}
  \epsilon_n u_n(i)&=\sum_{j}h_{ij}u_n(j)+\Delta(i) v_n(i) \\
  \epsilon_n v_n(i)&=-\sum_{j}h_{ij}^*v_n(j)+\Delta^*(i) u_n(i),
\end{aligned}
\end{equation}
where $h_{ij}=-t(\delta_{i,j+1}+\delta_{i,j-1})+[U(i)-\mu]\delta_{ij}$, $\Delta(i)$ is the superconducting pair potential, $\{\epsilon_n,\,u_n(i),\, v_n(i)\}$ are the energy and wavefunctions of quasiparticles with $n$ the quasiparticle quantum number (here the energy ordering number). The wavefunctions should be normalized, i.e. $\sum_i |u_n(i)|^2 + |v_n(i)|^2=1$, and satisfy the open boundary condition $u_n(0)=u_n(N+1)=0$ and $v_n(0) =v_n(N+1)=0$. The BdG Eqs. (\ref{bdg}) are numerically solved together with the self-consistency relation
\begin{equation}\label{op}
\Delta(i)=g\sum_{n} u_n(i) v_n^*(i) \big(1-2f_n\big),
\end{equation}
where $f_n=f(\epsilon_n)$ is the Fermi-Dirac quasiparticle distribution. The summation above includes positive-energy quasiparticle states inside the Debye window $0\leq\epsilon_n\leq\hbar\omega_D$, with $\omega_D$ the Debye frequency. Due to the time reversal symmetry, we regard $\Delta(i)$ as real.

For the half-filling the chemical potential $\mu$ is fixed by the relation
\begin{equation}
\bar{n}_e=1=\frac{1}{N}\sum_in_e(i),
\label{neav}
\end{equation}
where the electron occupation number $n_e(i)$ is given by
\begin{equation}\label{ne}
n_e(i) =\frac{2}{N} \sum_{n}\big[f_n |u_n(i)|^2+(1-f_n)|v_n(i)|^2\big].
\end{equation}

In our calculations we use the microscopic parameters $g=2$, $\hbar\omega_D=10$, and $N=301$. For this choice $\tau= (T_{cs}-T_{cb})/T_{cb} \approx 25\%$~(for zero field). However, as is mentioned above, $\tau$ can be higher for smaller values of the Debye frequency~\cite{bai2023}. Notice that $N=301$ is large enough to avoid any finite size effects. Generally, our qualitative conclusions are not influenced by this choice of the microscopic parameters. Below the energy-related quantities, the electric field and the temperature $T$ are shown in units of $t$, $t/(ea)$ and $t/k_B$, respectively. In our calculations, the self-consistent solution for $\Delta(i)$ is obtained with the accuracy of $10^{-8}$. 

\section{Results and discussions}\label{results}
\subsection{Suppression of $T_{cs}$ by electric fields}

\begin{figure}[!ht]
\centering
\includegraphics[width=1\linewidth]{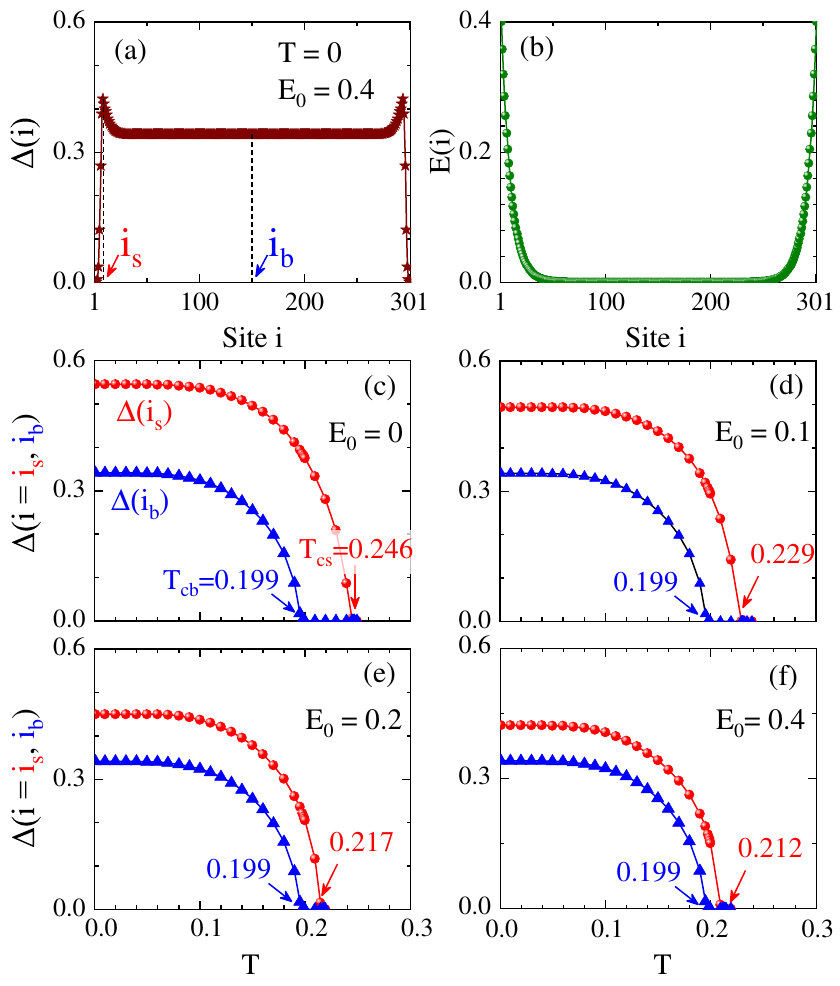}
\caption{(Color online) (a) The spatial profile of $\Delta(i)$ at $T=0$ and $E_0=0.4$; the site corresponding to the surface maximum is $i_s$ whereas $i_b=(N+1)/2$ marks the center of the chain. (b) The screened electric field versus $i$ for $E_0=0.4$. (c-f) $\Delta(i=i_s,\,i_b)$ versus $T$ at $E_0=0$, $0.1$, $0.2$ and $0.4$, respectively; $T_{cs}$ and $T_{cb}$ are defined as the critical temperatures at which $\Delta(i_s)$ and $\Delta(i_b)$ become zero.}
\label{fig1}
\end{figure}

In Fig.~\ref{fig1}(a) one can find a typical example of the self-consistent pair potential $\Delta(i)$ given as a function of the site number $i$. It is calculated for the electric-field strength $E_0=0.4$ at $T=0$. The spatial profile of $\Delta(i)$ demonstrates that the pair potential (the gap function or the order parameter) stays uniform inside the chain. It is close to $\Delta(i_b) =0.340$, where $i_b=(N+1)/2$. Below $\Delta(i_b)$ is regarded as the bulk pair potential. However, near the surface, the pair potential exhibits a maximum. Its value $\Delta_{max}=0.422$ is $24.1\%$ higher than $\Delta(i_b)$. The locus of the maximum is labeled as $i_s$, and here $i_s = 9$. 

The profile of the screened electric field is illustrated with $E_0=0.4$ in Fig.~\ref{fig1}(b). It vanishes in the region $i\in[50,\,250]$. Obviously, the maximum locus is in the domain of the exponential decay of the field. 

Figures~\ref{fig1}(c-f) show $\Delta(i_s)$ and $\Delta(i_b)$ as functions of $T$ for $E_0=0$, $0.1$, $0.2$ and $0.4$, respectively. The profiles of $\Delta(i_s)$ and $\Delta(i_b)$ are similar to the general temperature dependence of the BCS gap, however, each of these quantities drops to zero at a distinct critical temperature that depends $E_0$. As a result, we obtain the surface $T_{cs}$ and bulk $T_{cb}$ critical temperatures~\cite{samoilenka2020a,croitoru2020}. For $E_0=0$, we find $T_{cb}=0.198$ and $T_{cs}=0.246$ in agreement with the results reported in Ref.~\cite{bai2023}. We find that $T_{cb}$ is the same in Figs.~\ref{fig1}(c-f) and so, the electric field does not affect $T_{cb}$ due to the field screening. However, $T_{cs}$ is significantly affected by the field. For example, it decreases by about $14\%$ (from $0.246$ to $0.212$) as $E_0$ increases from $0$ to $0.4$.

\begin{figure}[t]
\centering
\includegraphics[width=0.7\linewidth]{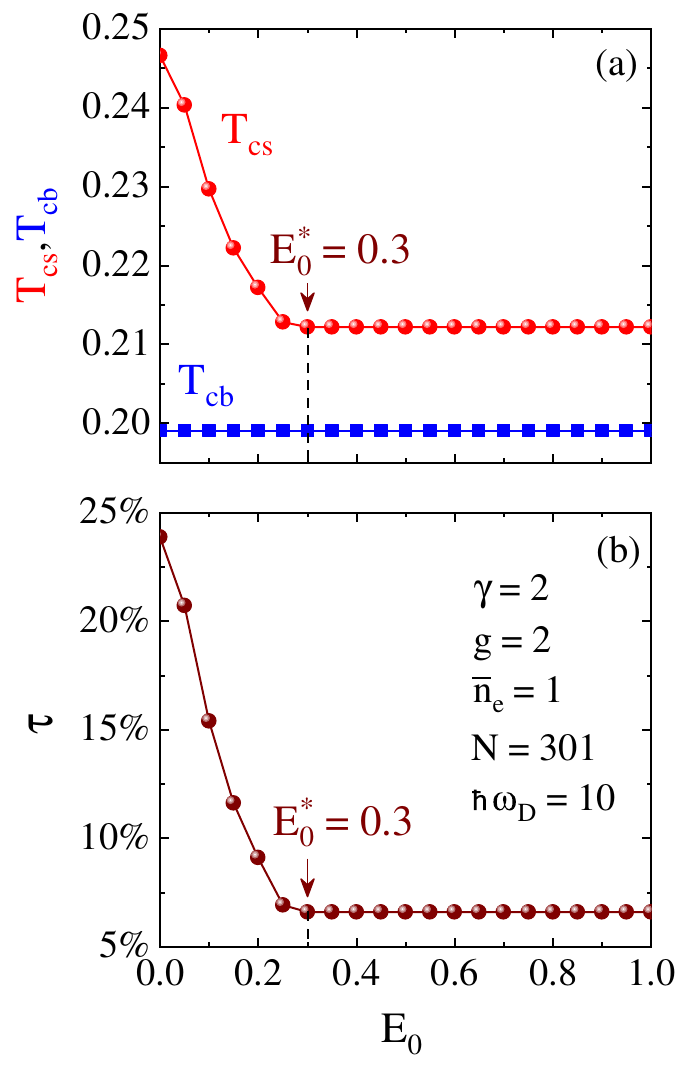}
\caption{(Color online) (a) The surface $T_{cs}$ and bulk $T_{cb}$ critical temperatures versus the field strength $E_0$. (b) The relative enhancement of the surface critical temperature $\tau=(T_{cs}-T_{cb})/T_{cb}$ as a function of $E_0$. $E_0^*$ denotes the field strength above which $T_{cs}$ does not change with $E_0$.}
\label{fig2}
\end{figure}

The critical temperatures $T_{cs}$, $T_{cb}$ and the relative enhancement of the surface superconducting temperature $\tau=(T_{cs}-T_{cb})/T_{cb}$ are shown in Figs.~\ref{fig2}(a,b). One can see that $T_{cb}$ stays constant ($T_{cb}=0.198$) when $E_0$ is varied in the range $0\leq E_0\leq 1$. Physically, this is clear as the screened electric field vanishes in the center of the chain at $i=i_b$, see Fig.~\ref{fig1}(b). Therefore, $\Delta(i_b)$ is not affected by the electric field together with the bulk critical temperature $T_{cb}$. As for $T_{cs}$, one finds that it decreases monotonically from $0.246$ at $E_0=0$ to $0.212$ at $E_0=E_0^*=0.3$. Further increasing $E_0$ does not have any effect on $T_{cs}$. It remains equal to $0.212$ when $E_0$ increases from $E_0^*$ to $1$, as seen from Fig.~\ref{fig2}(a). The corresponding relative enhancement of $T_{cs}$~[see Fig.~\ref{fig2}(b)] drops from $23.8\%$ at $E_0=0$ to $6.6\%$ at $E_0=E^*_0$ and then, stays the same for $E_0>E_0^*$. Thus, we find that the interference-induced surface superconductivity and its critical temperature can be fine-tuned by changing the applied electric field. For the chosen microscopic parameters this fine-tuning is within the range $\approx 7-24\%$. However, for smaller Debye frequencies the upper level of this range can increase up to $60-70\% $ at zero electric field.

\begin{figure}[!ht]
\centering
\includegraphics[width=1\linewidth]{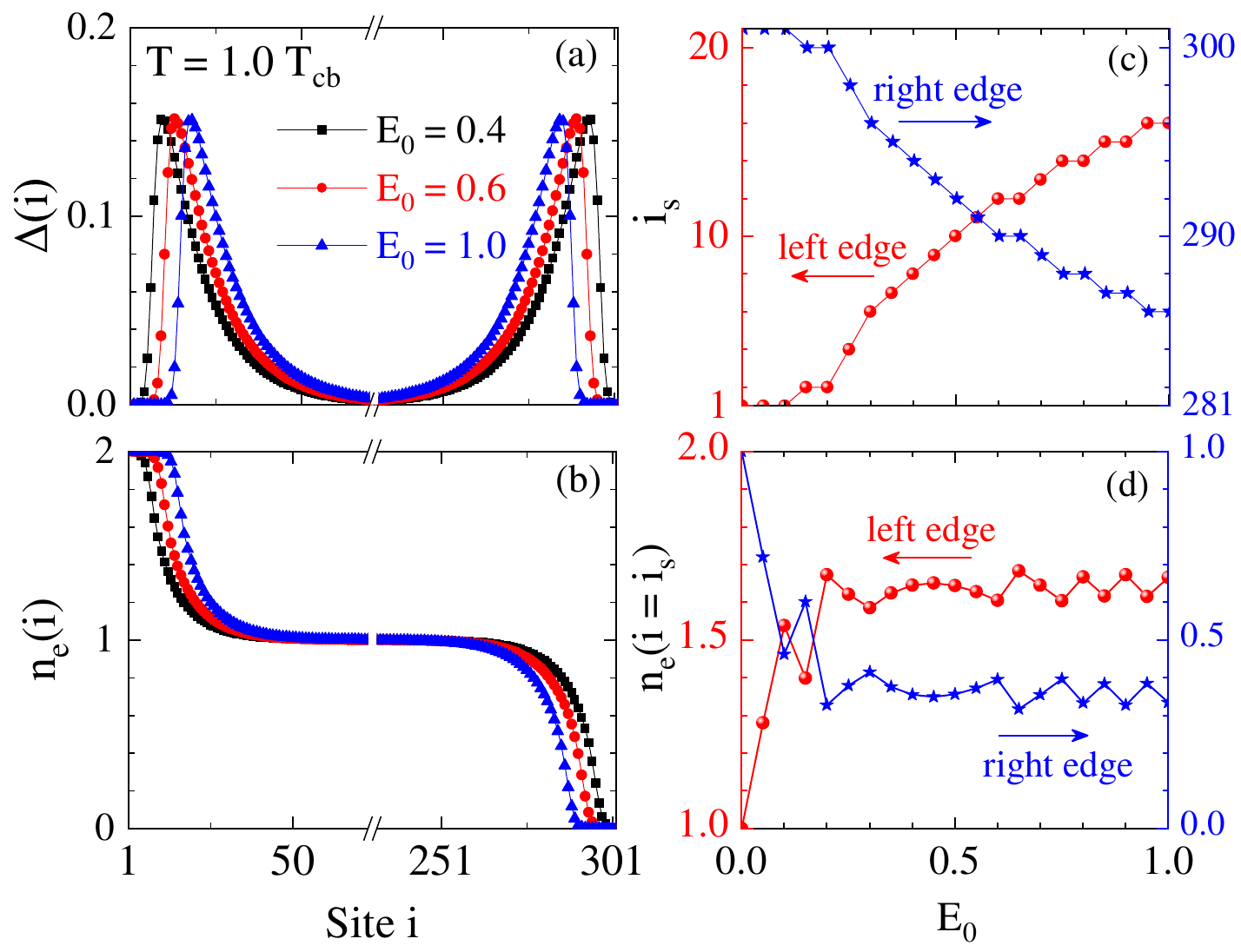}
\caption{(Color online) (a),(b) $\Delta(i)$ and $n_e(i)$ as functions of $i$, calculated at $T=1.0\,T_{cb}$ for $E_0=0.4$ (black squares), $0.6$ (red circles), and $1.0$ (blue triangles). (c) The locus $i_s$ of the pair-potential maximum near the right (blue stars) and left edges (red spheres) versus $E_0$. (d) The electron occupation number at $i=i_s$ near the right and left edges versus $E_0$, the colors and symbols are the same as in panel (c).}
\label{fig3}
\end{figure} 

We notice that the position of the pair-potential maximum, i.e. $i_s$, changes with $E_0$. This is seen from Fig.~\ref{fig3}(a), where the spatial profile of $\Delta(i)$ is shown for $T=1.0\,T_{cb}$ at $E_0=0.4$, $0.6$ and $1.0$. The surface enhancement is most pronounced at $i=i_s$ and the maximum shifts towards the chain center with increasing $E_0$. In more detail, there are two surface maxima, one is close to the left edge and situated at $i=i_s$, and another is located near the right edge. The both of them shift towards the chain center.  

In fact, the electric-field effect is even more complicated then one might expect from Fig.~\ref{fig3}(a). In Fig.~\ref{fig3}(b), one can see that the electron spatial distribution $n_e(i)$ is significantly altered in the presence of the field. While this quantity remains in the half-filling regime close to the center (in bulk), the electrons are pumped from the right edge and accumulate on the left. In particular, the sites near the right edge become completely empty while those near the left edge are fully occupied by electrons [$n_e(i)=2$], and the superconducting condensate vanishes for these sites. This indicates the emergence of the superconductor-insulator transition, which agrees with the findings of Ref.~\cite{yin2023}. Thus, we obtain two domains near the system edges: the first one (closer to the edges) is in the insulating state while the second domain exhibits an enhanced superconducting temperature in comparison with the bulk critical temperature. Since turning on an external electric field, we get the complex structure of different states - from the surface insulator to the surface-enhanced superconductor and further, to the bulk superconducting state.

\begin{figure}[!ht]
\centering
\includegraphics[width=0.6\linewidth]{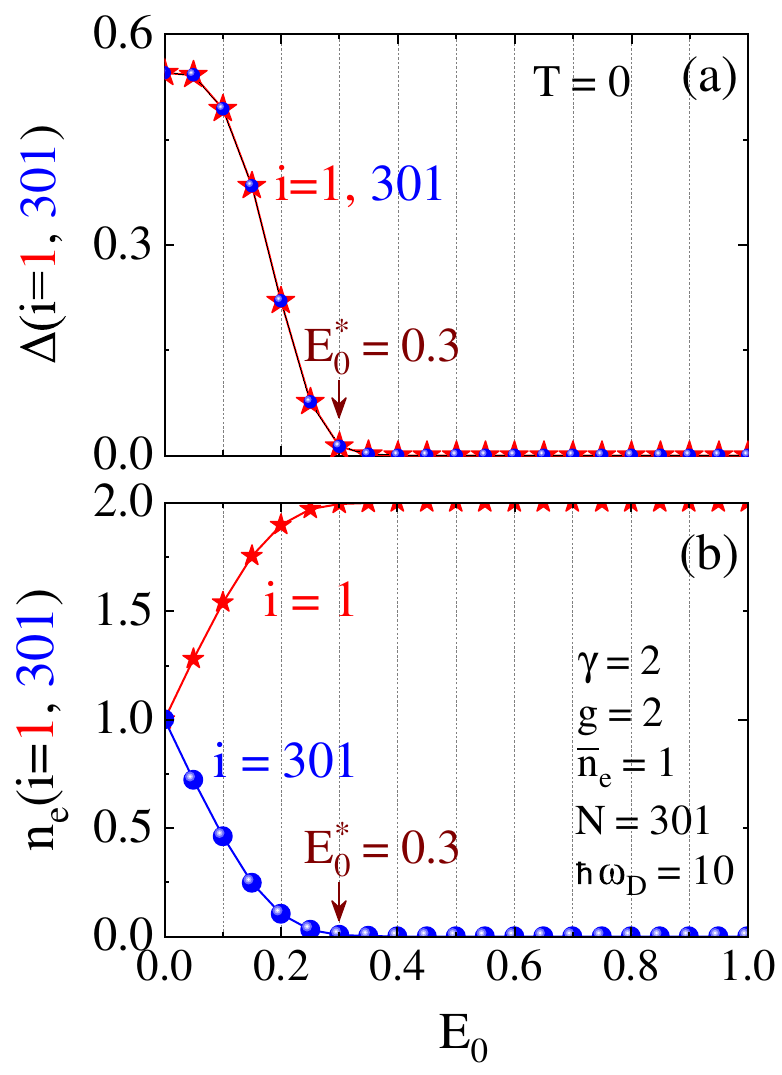}
\caption{(Color online) The pair potentials $\Delta(i)$~(a) and electron occupation number $n_e(i)$~(b) for $i=1$ and $301$ as functions of $E_0$ at $T=0$. The red stars correspond to $i=1$ while the blue spheres are for $i=301$. $E_0^*$ denotes the onset of the insulating state.}
\label{fig4}
\end{figure}

It is of importance to point out that the particular field strength $E^*_0$~(see Fig.~\ref{fig2}), above which $T_{cs}$ remains the same, marks actually the onset of the surface insulator state~\cite{yin2023}. This is seen from Fig.~\ref{fig4}(a,b), where the pair potential and electron occupation number at the first and last sites of the chain are shown versus $E_0$ at $T=0$. We obtain that the pair potentials at the first and last occupied sites become zero exactly at $E_0=E_0^*=0.3$, see panel (a). In turn, at the same value of the field strength (and above) we find $n_e(i=1)=2$ and $n_e(i=301)=0$. 

One can also see from Figs.~\ref{fig3}(a) and (b) that for $E_0>E_0^*$, the electron spatial distribution remains nearly the same in the vicinity of $i_s$. As a result, $\Delta(i_s)$, which is connected with $T_{cs}$, does not change with $E_0$ for $E_0>E_0^*$, which explains why $T_{cs}$ stays the same above $E_0^*$.

\subsection{Microscopic mechanism \\behind the suppression of surface superconductivity}

Now we investigate the microscopic mechanism underlying the suppression of the surface superconductivity induced by a screened electric field, based on the analysis of the quasiparticle contributions to the pair potential at $T=0$. To facilitate our study, we introduce the cumulative pair potential defined as~\cite{croitoru2020}
\begin{equation}\label{op2}
\Delta^{(\epsilon)}(i) = g\sum_{0\leq\epsilon_n\leq\epsilon} u_n(i) v_n^*(i) (1-2f_n).
\end{equation}
Below we consider the cumulative pair potential at $i=i_s$ and $i_b$. To simplify the notations, $\Delta^{(\epsilon)}(i=i_s)$ and $\Delta^{(\epsilon)}(i=i_b)$ are referred to as $\Delta^\epsilon_{s}$ and $\Delta^\epsilon_{b}$, respectively.

\begin{figure}[!ht]
\centering
\includegraphics[width=0.65\linewidth]{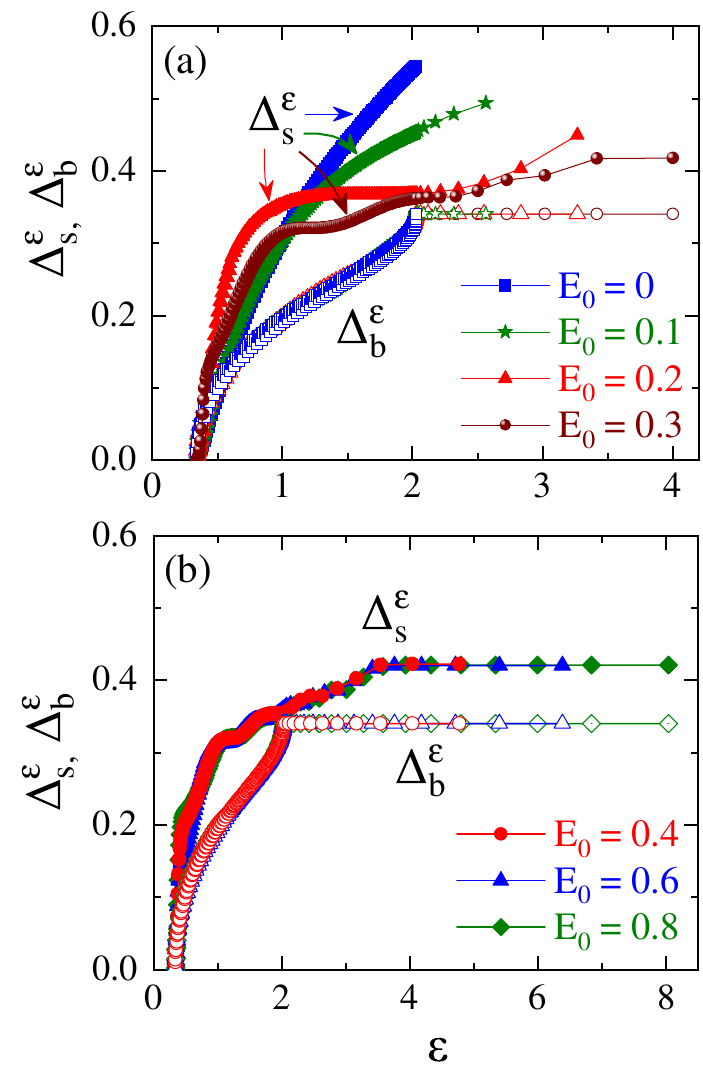}
\caption{(Color online) The cumulative pair potentials $\Delta_{s}^\epsilon \equiv \Delta^{(\epsilon)}(i_s)$ and $\Delta_{b}^\epsilon \equiv \Delta^{(\epsilon)}(i_b)$ versus the quasiparticle energy $\epsilon$ at $T=0$: panel (a) demonstrates the results for $E_0 = 0, \, 0.1,\, 0.2$, $0.3$; panel (b) shows the data for $E_0 = 0.4,\,0.6,\, 0.8$. The solid symbols correspond to $\Delta_{s}^{\epsilon}$ while the open ones are for $\Delta_{b}^{\epsilon}$.}
\label{fig5}
\end{figure}

Figure~\ref{fig5} demonstrates $\Delta_s^\epsilon$ and $\Delta_b^\epsilon$ as functions of the upper limit of the quasiparticle energy $\epsilon$ for the field strengths $E_0=0$, $0.1$, $0.2$, $0.3$, $0.4$, $0.6$, and $0.8$ at $T=0$. The results for $\Delta_s^\epsilon$ are shown by the solid symbols while those for $\Delta_b^\epsilon$ are given by the open symbols. As seen from Fig.~\ref{fig5}, all the quasiparticles have energies less than $\hbar\omega_D$ and so, every positive-energy quasiparticle state gives a contribution to the pair potential, according to Eq.~(\ref{op}). For $E_0=0$ [see the blue open squares in Fig.~\ref{fig5}(a)], one finds that the dependence of $\Delta_b^\epsilon$ on $\epsilon$ reflects the energy-dependence of the quasiparticle density of states (DOS) $D(\epsilon) = d \nu/d\epsilon$, with $d\nu$ the number of quasiparticles in the energy interval $d\epsilon$. $D(\epsilon)$ is proportional to the single-electron DOS $N(\xi)=d\nu/d\xi$, with $\xi$ the single-particle energy measured from the chemical potential $\mu$~($\mu=0$ for the half-filling case at zero field). Employing the simple BCS approximation $\epsilon=\sqrt{\xi^2+\Delta_0^2}$, with $\Delta_0$ the excitation gap~(the minimal quasiparticle energy), one finds $D(\epsilon)=N(\xi)\epsilon/\sqrt{\epsilon^2-\Delta_0^2}$. Due to the van Hove singularities at the lower and upper electron band edges, one obtains $N({\xi=\pm2})\rightarrow\infty$. In addition, $\epsilon/\sqrt{\epsilon^2-\Delta_0^2}\rightarrow\infty$, as $\epsilon$ approaches $\Delta_0$. Therefore, $\Delta_b^\epsilon$ has infinite derivatives at $\epsilon=\Delta_0\approx 0.3$-$0.4$ and $\epsilon=2$. 

When switching on the electrostatic field, we observe a similar dependence on $\epsilon$ for $\epsilon\leq2$, as seen from the data for $\Delta_b^\epsilon$ in Figs.~\ref{fig5}(a,b). However, for $E_0 > 0$ there appear high-energy quasiparticles with $\epsilon_n>2$ that do not produce any contribution to $\Delta_b^\epsilon$. This is seen from the flat profile $\Delta_b^\epsilon=0.34$ for $\epsilon >2$. Then, based on the Fig.~\ref{fig5}(a), we conclude that the bulk pair potential does not change with increasing $E_0$ though there are high-energy quasiparticles induced by the screened electric field. This conclusion is in agreement with our present results for $T_{cb}$ given in Fig.~\ref{fig2}(a).

The response of the cumulative pair potential at $i=i_s$ ($\Delta_s^\epsilon$) to the screened electric field is more complex. Here, when $E_0$ increases from $0$ to $0.1$, $\Delta_s^\epsilon$ remains nearly the same in the low-energy sector $\epsilon< 1.2$. However, its value decreases significantly as compared to that of $E_0=0$ for the energies $1.2<\epsilon < 2$. This decrease is partly compensated by the appearance of the quasiparticle contributions with $\epsilon >2$. The dependence of $\Delta_s^\epsilon$ on $\epsilon$ demonstrates further evolution at $E_0=0.2$. Its overall increase with $\epsilon$ becomes more pronounced for $\epsilon <1.2$, as compared to the case of $E_0=0.1$. Then, $\Delta_s^\epsilon$ stays nearly flat for $1.2<\epsilon<2$, with $\Delta_s^\epsilon \approx 0.367$, while it slightly increases with $\epsilon$ for the high-energy regime with $\epsilon >2$. For $E_0=0.3$, the spatial profile of $\Delta_s^\epsilon$ becomes even more complex. One can see the presence of three flat regions around the points $\epsilon=0.2$, $2$, and $3.5$. Quasiparticles with the corresponding energies do not contribute to the surface superconducting state. 

Finally, for $E_0 > 0.3$ the results for $\Delta_s^\epsilon$ does not change any more, which is in agreement with our finding that $T_{cs}$ does not change with $E_0$ for $E_0>E_0^*=0.3$, see Fig.~\ref{fig2}. One can see from Fig.~\ref{fig5}(b) that in this regime $\Delta_s^\epsilon$ exhibits a faster overall increase with $\epsilon$ for low energies, as compared to the corresponding increase of $\Delta_b^\epsilon$. The surface cumulative pair potential reaches the values $0.42$ at $\epsilon = 4.0$ and stays the same for $\epsilon > 4.0$. A similar high-energy behavior is obtained for the bulk cumulative pair potential. However, it saturates at the smaller value $0.34$ when $\epsilon$ exceeds $2.0$. This is in agreement with the fact that for $E_0>E^*_0$ we find $T_{cs}$ larger than $T_{cb}$ by $6.6\%$. Thus, our study demonstrates that the alterations of the quasiparticle contributions with $\epsilon > 1.2$ are responsible for the changes of the surface states in the presence of the external electric field. 

\begin{figure}[ht]
\centering
\includegraphics[width=1\linewidth]{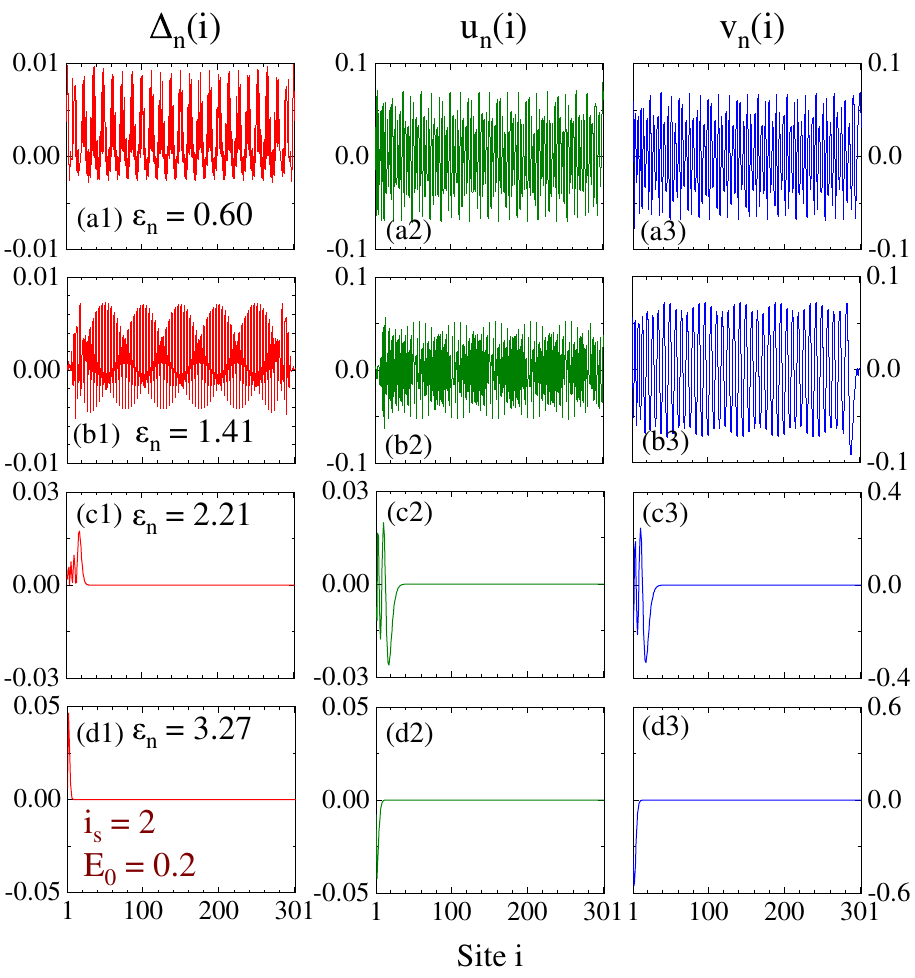}
\caption{(Color online) The single-species quasiparticle contribution $\Delta_n(i)$ and the quasiparticle wave functions $u_n(i)$ and $v_n(i)$ for $\epsilon_n=0.60$ (a), $1.41$ (b), $2.21$ (c) and $3.27$ (d), as calculated at $E_0=0.2$ and $T=0$. Panels (a1-d1) demonstrate $\Delta_n(i)$; panels (a2-d2) and (a3-d3) give the corresponding $u_n(i)$ and $v_n(i)$, respectively. In this case, the maximum of $\Delta(i)$ is located at $i_s=2$.}
\label{fig6}
\end{figure}

\begin{figure}[!ht]
\centering
\includegraphics[width=1\linewidth]{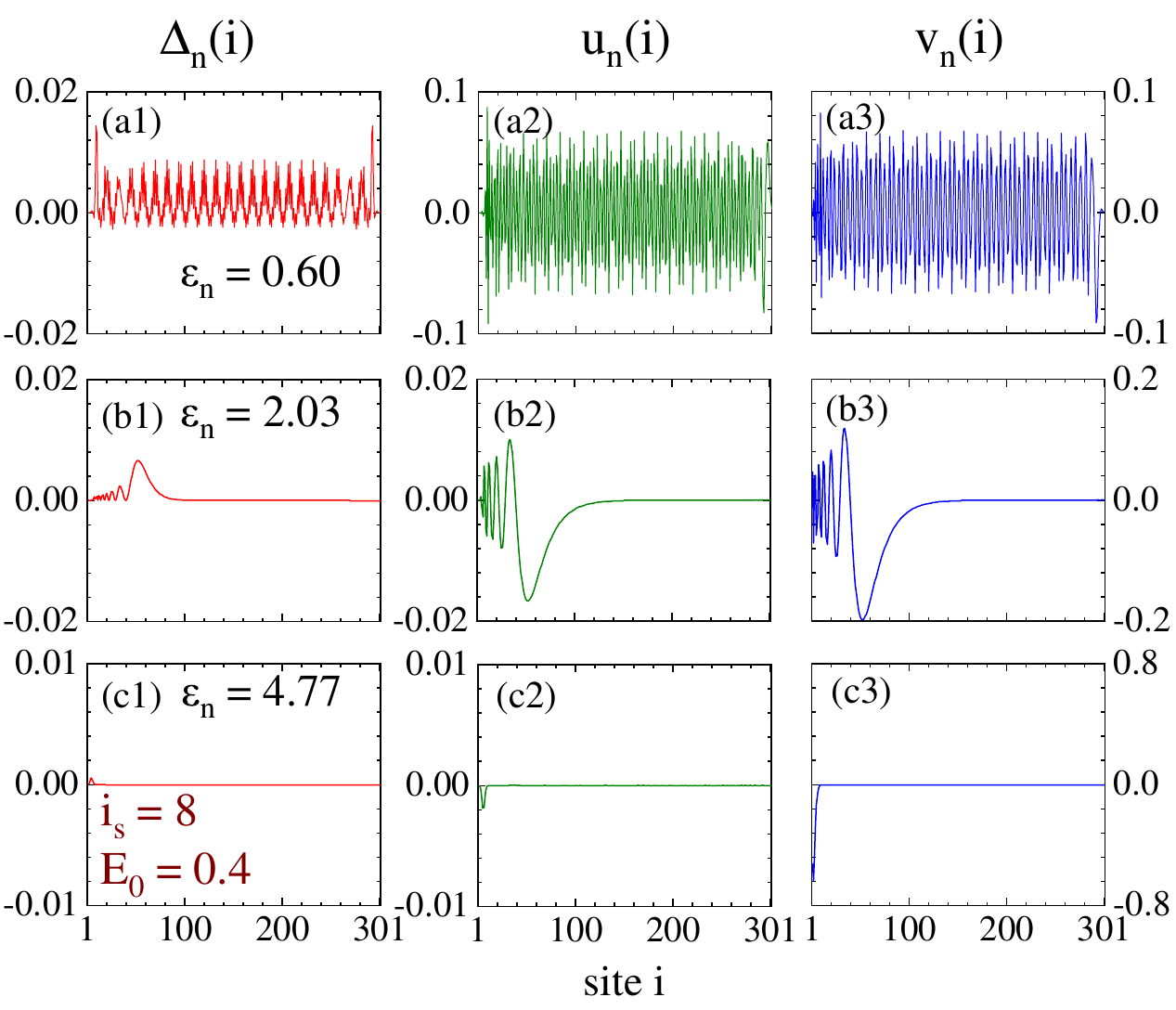}
\caption{(Color online) The same as in Fig.~\ref{fig6} but for $\epsilon_n=0.60$, $2.03$ and $4.77$ and $E_0 = 0.4$. Here we have $i_s=8$.}
\label{fig7}
\end{figure} 

A further insight is obtained when analysing the single-species quasiparticle contribution to the pair potential given by
\begin{equation}
\Delta_n(i) = g u_n(i) v_n^*(i) (1-2f_n). 
\end{equation}
Figure~\ref{fig6} shows $\Delta_n(i)$, $u_n(i)$ and $v_n(i)$ for four quasiparticle species with $\epsilon_n=0.6$~[panels (a1-a3)], $1.41$~[panels (b1-b3)], $2.21$~[panels (c1-c3)], and $3.27$~[panels (d1-d3)]. The results are obtained for $E_0=0.2 < E^*_0$. In this case, the left maximum of $\Delta(i)$ is located at $i_s=2$. For $\epsilon_n=0.6$, see Figs.~\ref{fig6}(a1-a3), $\Delta_n(i)$ is a strongly oscillating function of $i$, together with $u_n(i)$ and $v_n(i)$. Here we find that $\Delta_n(i_s)=0.01$~(it reaches its local maximum) whereas $\Delta_n(i_b)=0.009$. This highlights the fact that the low-energy quasiparticles give almost the same contribution to the surface and bulk superconductivity for sufficiently small fields, i.e. the screened electric field does not significantly affect the contributions of these quasiparticles to the pair potential. For $\epsilon_n=1.41$, see Figs.~\ref{fig6}(b1-b3), the surface-superconductivity contribution is nearly suppressed. Indeed, we have $\Delta_n(i_s) = 7.6\times10^{-5}$, as compared to $\Delta_n(i_b)= 7.1\times 10^{-3}$. Such a small value of $\Delta_n(i_s)$ corresponds to the first flat regime of $\Delta_s^\epsilon$ around the energy $\epsilon \approx1.2$ in Fig.~\ref{fig5}. At the same time $\Delta_b^\epsilon$ is still significant. 

When $\epsilon_n$ exceeds $2$, the corresponding quasiparticles do not contribute to the bulk superconductivity, i.e. $\Delta_n(i_b)$ becomes negligible, as seen from the examples with $\epsilon_n=2.21$ and $3.27$ shown in Figs.~\ref{fig6}(c1-c3) and (d1-d3). However, for the surface contribution we have $\Delta_n(i_s)= 1.9\times10^{-3}$~(for $\epsilon_n=2.21$) and $4.6\times10^{-2}$~(for $\epsilon_n=3.27$). The wave functions $u_n(i)$ and $v_n(i)$ for quasiparticles with $\epsilon_n>2$ are localized near the chain edges due to the presence of the screened electric field, see also Ref.~\cite{yin2023}. 

Figure~\ref{fig7} shows $\Delta_n(i)$, $u_n(i)$ and $v_n(i)$ for the three quasiparticle species with $\epsilon_n=0.6$~[panels (a1-a3)], $2.03$~[panels (b1-b3)], and $4.77$~[panels (c1-c3)]. The calculations are performed for $E_0=0.4 > E^*_0$. Notice that for this case $i_s=8$. For $\epsilon_n=0.6$, the general behavior of $\Delta_n(i)$, $u_n(i)$, and $v_n(i)$ is similar to that of Figs.~\ref{fig6}(a1-a3). The results calculated for $\epsilon_n=2.03$ and demonstrated in Figs.~\ref{fig7}(b1-b3) are similar to those in Figs.~\ref{fig6}(c1-c3). Finally, the data shown in Figs.~\ref{fig7}(c1-c3) do not have a similar data-set in Fig.~\ref{fig6}. The point is that Figs.~\ref{fig7}(c1-c3) correspond to the quasiparticle species which produces a negligible contribution to the bulk pair potential [$\Delta_n(i_b)=4.3\times10^{-5}$]. At the same time, its surface contribution is also strongly suppressed. These high-energy quasiparticle species corresponds to the long nearly flat regime of $\Delta_s^\epsilon$ illustrated in Fig.~\ref{fig5}(b). 
     
\section{Conclusions}
\label{conclusions}

In summary, we have investigated the effect of an external electrostatic field on the interference-induced surface superconductivity. Our study is based on a self-consistent solution of the Bogoliubov-de Gennes equations for the one-dimensional attractive Hubbard model with the nearest-neighbor hopping and half-filling level. To simplify our consideration, a phenomenological expression has been introduced for the screened electric field. Our results demonstrate that the surface critical temperature $T_{cs}$ is sensitive to the electric field so that the surface superconductivity can be tailored by changing the field strength. It is worth noting that the field shifts the surface maxima of the superconductive pair potential towards the center of the system so that one gets the combination of the surface insulating (closer to the edges) and surface superconducting (further from the edges) domains. When the field strength exceeds its critical value, the surface superconducting temperature does not change any more. In this case increasing $E_0$ is only accompanied by a further shift of the surface pair-potential maxima towards the chain center. The corresponding maximal value of the pair potential and $T_{cs}$ are not altered. 

\section*{ACKNOWLEDGMENTS}
This work was supported by Zhejiang Provincial Natural Science Foundation (Grant No. LY18A040002) and Science Foundation of Zhejiang Sci-Tech University(ZSTU) (Grant No. 19062463-Y). The study has also been funded within the framework of the HSE University Basic Research Program.


\end{document}